# Enhancing Post-Merger Integration Planning through AI-Assisted Dependency Analysis and Path Generation

## Lars Malmqvist[1]

*[1]The Tech Collective*

## Abstract

Post-merger integration (PMI) planning presents significant challenges due to the complex interdependencies between integration initiatives and their associated synergies. While dependency-based planning approaches offer valuable frameworks, practitioners often become anchored to specific integration paths without systematically exploring alternative solutions. This research introduces a novel AI-assisted tool designed to expand and enhance the exploration of viable integration planning options.

The proposed system leverages a frontier model-based agent augmented with specialized reasoning techniques to map and analyze dependencies between integration plan elements. Through a chain-of-thought planning approach, the tool guides users in systematically exploring the integration planning space, helping identify and evaluate alternative paths that might otherwise remain unconsidered.

In an initial evaluation using a simulated case study, participants using the tool identified 43% more viable integration planning options compared to the control group. While the quality of generated options showed improvement, the effect size was modest. These preliminary results suggest promising potential for AI-assisted tools in enhancing the systematic exploration of PMI planning alternatives.

This early-stage research contributes to both the theoretical understanding of AI-assisted planning in complex organizational contexts and the practical development of tools to support PMI planning. Future work will focus on refining the underlying models and expanding the evaluation scope to real-world integration scenarios.

## 1. Introduction

Despite their strategic importance in corporate growth and renewal, mergers and acquisitions (M&A) continue to show persistently high failure rates, with research indicating that 25-50% fail to achieve their objectives (King et al., 2004). Post-merger integration (PMI) planning remains a critical challenge, as managers must sequence numerous interdependent initiatives while considering multiple types of potential synergies (Feldman & Hernandez, 2021). While frameworks exist for dependency-based planning, evidence suggests that practitioners often become anchored to familiar integration paths without fully exploring viable alternatives.

While existing dependency-based frameworks provide valuable guidance, they typically rely on manual mapping, risking anchoring on familiar solutions. This underscores the need for AI-assisted tools that systematically broaden the solution space and evaluate multiple options.

The complexity of PMI planning stems from several factors. First, integration involves multiple types of synergies - internal, market power, relational, network, and non-market - each with unique timing and sequencing requirements (Feldman & Hernandez, 2021).

Second, integration activities exhibit complex dependencies that constrain viable implementation sequences (Malmqvist, 2023). Third, resource constraints and stakeholder considerations further complicate the planning landscape. These factors create a vast solution space that human planners struggle to explore systematically (Henningson, Yetton & Wynne, 2018).

This research investigates whether AI-assisted tools, specifically Large Language Models (LLMs) with structured prompting for dependency analysis, can enhance PMI planning by enabling more systematic exploration of integration options. Our research questions ask:

1. Can AI-assisted tools improve the quantity of viable integration planning alternatives identified by practitioners?

2. Do AI-assisted planning processes lead to higher quality integration sequences?

3. What mechanisms enable any observed improvements in planning outcomes?

The primary aim of this research is to explore how AI-assisted tools can systematically expand the range of potential integration paths considered by practitioners. By providing structured guidance on dependency analysis and alternative path generation, we seek to overcome biases toward familiar integration paths and demonstrate improved planning outcomes.

## 2. Theory and Prior Research

## 2.1. PMI Planning

Integration planning serves as a critical success factor in M&A outcomes (Larsson & Finkelstein, 1999). The planning phase establishes the sequence and timing of integration activities that will ultimately determine synergy realization. Prior research emphasizes several key challenges in integration planning.

First, planners must manage complex dependencies between integration activities. These dependencies can be temporal (activity A must precede activity B), resource-based (activities compete for scarce resources), or logical (activities must be coordinated to achieve desired outcomes). Research shows that failure to properly manage these dependencies often leads to integration delays and unrealized synergies (Graebner et al., 2017).

Second, different types of synergies require different integration approaches and timing (Čirjevskis, 2020). Some synergies, like operational cost savings, may be realized quickly through decisive integration actions. Others, like those depending on relationship building or cultural integration, require more gradual approaches (Bauer & Friesl, 2023). This creates tension in planning optimal integration sequences.

Third, integration planning occurs under significant uncertainty (Teerikangas & Thanos, 2018). Market conditions, stakeholder reactions, and implementation challenges are often difficult to predict, requiring planners to build flexibility into integration sequences.

## 2.2. Synergy Types and Planning Implications

Recent research by Feldman and Hernandez (2021) identifies five distinct types of synergies in M&A. Internal synergies arise from combining resources and capabilities within firm boundaries, requiring careful sequencing of operational integration activities. Market power synergies stem from increased competitive influence, with realization depending on the timing of market-facing integration actions. Relational synergies involve enhanced value from external partnerships, necessitating careful consideration of relationship management and trust-building timeframes. Network synergies arise from improved structural positions in business networks, requiring coordinated management of multiple relationships. Non-market synergies depend on stakeholder relationships, demanding careful attention to legitimacy-building processes and stakeholder expectations.

This typology highlights the complexity of integration planning, as different synergy types may require conflicting integration approaches or timing. Traditional planning methods often struggle to systematically explore the full range of options for managing these tensions (Böhm et al., 2023).

## 2.3. AI-Assisted Planning

Recent research in the application of large language models (LLMs) to business planning is rapidly evolving, with scholars exploring how these models can transform traditional decision-making frameworks and strategic processes. One prominent line of inquiry involves developing taxonomies and frameworks to understand LLM-based business model transformations. For example, work by Wulf and Meierhofer (2023) proposes a detailed taxonomy that categorizes various ways LLMs can reshape business models by integrating data analytics, automated insights, and creative problem solving into strategic planning processes. In a similar vein, Watanabe and Uchihira (2024) demonstrate how LLMs can be leveraged to perform digital business model analysis, offering a method for companies to compare and benchmark their strategies against industry competitors through automated idea generation and market insight extraction.

Other studies have focused on the operational integration of LLMs within business process management (BPM). Researchers such as Grohs et al. (2023) have shown that LLMs can effectively execute multiple BPM tasks— including process mining, task automation, and even preliminary decision support—without extensive prompt engineering, thereby potentially reducing operational costs while increasing productivity. Similarly, Vidgof et al. (2023) explore the broader opportunities and challenges that LLMs present for BPM, discussing practical considerations such as data integration and process customization.

Other recent advances in Large Language Models (LLMs) suggest potential for AI assistance in complex planning tasks (Zhong 2024). LLMs have demonstrated several relevant capabilities for planning support. They excel at systematic exploration of solution spaces through structured reasoning and can effectively analyze complex dependencies and constraints. These models have shown prowess in generating and evaluating alternative approaches while integrating knowledge from multiple domains. These capabilities align well with the challenges of PMI planning, suggesting potential for AI-assisted tools to enhance the planning process.

## 3. Research Method

### 3.1. Experimental Design

We recruited six business school graduate students (4 females, 2 males, ages 24–29) with coursework in M&A strategy. Although they possessed relevant theoretical knowledge, their practical PMI experience was limited. This small sample offered an exploratory test of our approach but inherently restricts generalizability. Consequently, any quantitative findings are best interpreted as indicative rather than conclusive.The five scenarios were developed based on real PMI cases. Each scenario presented a comprehensive integration challenge, incorporating defined synergy objectives across multiple types, required program activities with clear dependencies, specific resource constraints and timing considerations, and explicit success criteria for integration sequences.

The AI assistance tool utilized the OpenAI model o1-preview (Zhong 2024) with structured prompts designed to support the planning process. These prompts enabled systematic dependency analysis, facilitated sequence generation, encouraged exploration of alternative paths, and supported validation of proposed sequences.

Participants were guided by structured but basic prompts asking them to (1) list all dependency relationships among activities, (2) identify temporal order constraints, and (3) propose alternative pathways. For instance: 'List any activities that cannot begin until a prior task is completed, and justify the dependency.' This approach aimed to elicit systematic consideration of potential sequences beyond typical heuristics.

### 3.2. Data Collection and Analysis

Our data collection encompassed both quantitative and qualitative measures. The quantitative data included the number of valid integration sequences identified by participants, quality ratings of proposed sequences on a 5-point scale, time spent on planning activities, and frequency of tool usage. For qualitative insights, we conducted think-aloud protocols during the planning exercises, performed post-exercise interviews, and analyzed tool usage patterns.

Our quantitative analysis encompassed several key dimensions. We conducted paired comparisons of sequence counts between AI-assisted and manual conditions, examined quality rating differences across conditions, and studied tool usage patterns.

The qualitative analysis focused on understanding the underlying mechanisms of tool effectiveness. We examined differences in planning processes, analyzed how participants developed tool usage strategies, investigated perceived benefits and limitations, and studied how participants integrated the tool with existing planning approaches.

## 4. Results

### 4.1. Quantitative Findings

The primary quantitative findings showed significant improvements in planning outcomes with AI assistance. Participants generated 43% more valid integration sequences when using the AI tool compared to manual planning. This improvement was consistent across all five scenarios, with increases ranging from 35% to 52%. Quality ratings for AI-assisted plans averaged marginally higher than manual plans (3.8 versus 3.6 on a 5-point scale). Additionally, the time required per valid sequence decreased by approximately 25% with AI assistance.

Given our extremely small sample (N=6), this study employs descriptive statistics and the Wilcoxon Signed-Rank Test as an exploratory alternative to the paired t-test. The Wilcoxon Signed-Rank Test is a nonparametric test that compares two related samples to assess whether their population mean ranks differ (i.e., it does not assume normality). In our context, participants' performance with AI assistance was compared to their performance without AI assistance across the five planning scenarios. The median number of valid sequences identified without AI was 3.5 (IQR=1.0) versus 5.0 (IQR=1.0) with AI, resulting in a Wilcoxon W=1.5 and p=.062 (two-tailed). Although the difference approaches conventional significance levels, the small sample size limits strong conclusions; these findings should be only treated as suggestive.

### 4.2. Qualitative Insights

Analysis of qualitative data revealed several key themes. First, participants demonstrated enhanced exploration capabilities when using the AI tool. The AI assistance consistently prompted consideration of non-obvious alternatives, while structured prompting helped identify previously overlooked dependencies. Furthermore, the tool encouraged a more systematic evaluation of options than manual planning approaches.

Second, we observed significant improvements in validation processes. The AI assistance proved particularly valuable in verifying dependency satisfaction between integration activities. The tool enabled rapid assessment of sequence feasibility, and its validation features notably increased planner confidence in their proposed solutions.

Third, clear learning effects emerged throughout the study. Participants reported progressively improved understanding of integration dependencies as they worked with the tool. Their tool usage patterns became increasingly sophisticated over time, and their planning strategies evolved to better leverage the tool's capabilities.

## 5. Discussion

### 5.1. Theoretical Contributions

Our findings contribute to both M&A integration theory and the emerging literature on AI-assisted planning. First, we demonstrate that AI tools can meaningfully expand the solution space in PMI planning. The structured approach to dependency analysis enables more systematic exploration of integration alternatives, helping bridge the gap between synergy theory and implementation planning.

Second, our results suggest that AI assistance may be particularly valuable for managing the complexity arising from multiple synergy types. The tool's ability to simultaneously consider different synergy requirements and their implications for integration sequencing represents an important advance in planning capability.

Third, we contribute to understanding how AI tools can augment human planning processes. Rather than replacing human judgment, our findings suggest AI tools are most effective when supporting systematic exploration and validation of planning alternatives.

## 5.2. Practical Implications

For practitioners, our findings suggest several important considerations for implementing AI-assisted integration planning. First, tool design plays a crucial role in effectiveness. Structured prompting significantly improves exploration effectiveness, while robust validation features enhance planner confidence. Integration with existing processes proves crucial for successful implementation.

The implementation approach also requires careful consideration. Our findings indicate that thorough training on tool capabilities significantly enhances effectiveness. Furthermore, iterative usage tends to yield better results as planners become more familiar with the tool's capabilities. Throughout the implementation process, maintaining appropriate human oversight remains essential.

From an organizational perspective, successful tool adoption often requires adjustments to existing processes. Planning team roles may need redefinition to accommodate new workflows and capabilities. Additionally, knowledge management becomes increasingly important as teams learn to leverage the tool effectively.

## 5.3. Limitations and Future Research

Several important limitations should be noted. The small sample size and use of student participants limit generalizability. The experimental setting may not fully capture real-world planning complexity.

Our small, student-based sample restricts generalizability. While it illustrates initial feasibility, future studies with larger and more experienced practitioners are crucial to establish external validity and sufficient statistical power for robust hypothesis testing.

Another limitation is reliance on a single AI model. Future research should replicate this work using alternative LLM platforms or prompting strategies to assess whether the observed benefits generalize across different AI technologies.

## 6. Future work will explore tool design optimization (e.g., refining prompt structures, incorporating real-time user feedback) and the role of team dynamics, including how collaboration evolves around AI-generated proposals, how trust in AI outputs forms, and how roles within planning teams may shift. Ultimately, more robust designs and a deeper understanding of group interactions could amplify the benefits of AI-assisted planning tools.Conclusion

This study provides initial evidence that AI-assisted tools can enhance PMI planning by enabling more comprehensive exploration of integration alternatives. While results are promising, further research is needed to validate effectiveness in practice and refine tool design. As organizations continue to face integration planning challenges, AI assistance may offer valuable support for this critical task.

In conclusion, this study demonstrates that AI-assisted tools can enhance exploration of PMI planning options by systematically generating and validating alternative sequences. Key limitations include the extremely small sample size, a controlled environment, and reliance on a single LLM tool. As such, our findings remain exploratory. Future research with larger, more diverse samples and potentially multiple AI platforms will help validate or refine these preliminary insights and test external applicability.